\documentclass{article}

\usepackage[utf8]{inputenc}
\usepackage{tismir}
\usepackage{amsmath}
\usepackage{hyperref}
\usepackage{url}
\usepackage{graphicx}
\usepackage{booktabs}
\usepackage{lipsum}

\title{Generative AI Training and Copyright Law}
\author{%
Sebastian Stober\thanks{Artificial Intelligence Lab, Faculty of Computer Science, Otto-von-Guericke-University Magdeburg, Universitätsplatz 2, 39106 Magdeburg, Germany}
\ ~and %
Tim W.~Dornis\thanks{Institute of Legal Informatics, Faculty of Law, Leibniz University Hannover, Königsworther Platz 1, 30167 Hannover, Germany}%
}

\date{}

\type{overview}

\authorref{Stober,~S. and Dornis,~T.~W.}
\authorshort{Stober and Dornis}
\titleshort{Generative AI Training and Copyright Law}

\begin{document}


\twocolumn[{%
\maketitleblock
\begin{abstract}
Training generative AI models requires extensive amounts of data. 
A common practice is to collect such data through web scraping. 
Yet, much of what has been and is collected is copyright protected. 
Its use may thus be copyright infringement.
In the USA, AI developers rely on ``fair use'' and in Europe, the prevailing view is that the exception for ``Text and Data Mining'' (TDM) applies. 
In a 2024 interdisciplinary tandem-study, we have argued in detail that this is actually not the case because generative AI training fundamentally differs from TDM. 
In this article, we share our main findings and the implications for both public and corporate research on generative models. 
We further discuss how the phenomenon of training data memorization leads to copyright issues independently from the ``fair use'' and TDM exceptions. 
Finally, we outline how the MIR community could contribute to the ongoing discussion about fair practices with respect to generative AI that satisfy all stakeholders.
\end{abstract}
\begin{keywords}
Generative Training, 
Copyright,
Memorization,
Data Provenance
\end{keywords}
}]
\saythanks{}



\section{Introduction}\label{sec:headings}

Generative artificial intelligence (AI) has emerged as a transformative tool across numerous creative domains, including music, visual art, and literature. 
These systems operate by learning patterns, structures, and styles from large datasets, enabling them to generate novel outputs that mimic or extend human creativity. 
Recent advances, particularly in deep learning architectures such as transformer models and diffusion-based approaches, have significantly enhanced the quality and diversity of AI-generated music. 
Models like Jukebox \citep{jukebox} or MusicLM \citep{musiclm} 
as well as text-to-music AI services such as provided by the start-ups Suno AI\footnote{\href{https://suno.com}{https://suno.com}.} and Udio\footnote{\href{https://www.udio.com/}{https://www.udio.com/}.} exemplify the state of the art, producing compositions that range from stylistic imitations to innovative genre-blending works.

A key factor underlying this progress is the scale and richness of the training data. 
High-quality, diverse datasets allow models to capture the intricacies of musical styles, genres, and techniques. 
However, the reliance on extensive copyrighted material raises complex legal and ethical questions \citep{Goetze2024}. 
For instance, in a legal statement\footnote{United States District Court for the District of Massachusetts, Civil Action No.~1:24-cv-11611-FDS, Document 28, Filed 08/01/24. Archived at \href{https://archive.org/details/gov.uscourts.mad.272063/gov.uscourts.mad.272063.28.0.pdf}{https://archive.org/details/gov.uscourts.mad.272063/ gov.uscourts.mad.272063.28.0.pdf}.} as part of a copyright lawsuit by the Record Industry Association of America (RIAA) against Suno AI, the company openly stated that ``Suno's training data includes essentially all music files of reasonable quality that are accessible on the open Internet, abiding by paywalls, password protections, and the like, combined with similarly available text descriptions.''
They further argued that ``It is fair use under copyright law to make a copy of a protected work as part of a back-end technological process, invisible to the public, in the service of creating an ultimately non-infringing new product." 
Time will tell whether this argument will hold in court.
Regardless of the outcome, this case illustrates how it has become increasingly difficult to navigate the legal space in the context of generative AI.

Our analysis builds upon a rich interdisciplinary discourse.
The legal and philosophical questions raised by generative AI intersect with a longer-standing critical discourse on musical property and authorship. 
\citet{Scherzinger2014} 
  critiques the historical expansion of intellectual property rights in music, arguing that increasingly broad copyright protections contribute to the ``enclosure'' of shared cultural resources. 
  Because musical creativity relies heavily on borrowing, repetition, and transformation, treating musical elements as exclusive property can paradoxically inhibit the innovation and expressive freedom that copyright is meant to encourage. 
This tension between protection and progress is central to the current debate on AI training data. 
\citet{Drott2019} 
  applies a similar line of analysis to the music AI industry, examining the political economy of how corporate AI development extracts value from musical works used as training data and raising fundamental questions about copyright, fair compensation, and the sustainability of a musical commons.
  Taking a philosophical approach to the question of AI-generated output, \citet{Caldwell2023} examines who (or what) should be considered the ``author'' of AI art. 
  Drawing on the theories of Locke, Hegel, and Kant, she argues that the end user -- whose creative choices guide the system and shape the resulting work -- should be treated as the author.
  Her position challenges inter alia the current U.S. Copyright Office's policy denying protection to works lacking sufficient human authorship.
  This debate over authorship has practical consequences for documentation and cataloging, as highlighted by \citet{Berkowitz2024}. 
  His analysis of recent U.S. Copyright Office and library cataloging guidelines reveals them 
  internally inconsistent and poorly suited to describing human–AI co-creation, and he advocates for more nuanced cataloging practices oriented toward accurately representing the creative process.
  Finally, \citet{Clancy2021}  provides a comprehensive systemic perspective in his doctoral thesis, modeling the broader ``music ecosystem'' affected by AI technologies. 
  He argues that existing copyright frameworks are insufficient to address the economic disruptions introduced by AI-generated music and that sustainable responses will require not only legal reform but also 
  a foundational ethical response from all stakeholders.
Collectively, these works establish that the challenges posed by generative AI are not merely technical or doctrinal, but are deeply embedded in ongoing struggles over musical value, authorial identity, and the ethical governance of cultural production -- a context that underscores the need for the pragmatic documentation framework we propose.

With this paper, we aim to create a better understanding of the interplay between generative AI's technical capabilities and copyright law 
-- particularly in the context of music information retrieval (MIR) and its broader implications for creators, researchers, and industry stakeholders.
It is based on a study on generative AI and copyright law in Germany and the EU conducted by the authors in 2024 and 
published in German as an open-access book \citep{2024:dornis:urheberrec}.
This article provides a summary of the main findings with specific implications for the MIR community. 
Section \ref{sec:legal_frameworks} introduces legal frameworks for training data collection and usage and discusses their applicability to training generative AI models.
The text extends the scope of the original study by incorporating a discussion of the fair use doctrine in the US and a brief coverage of regulations in jurisdictions other than the EU and the US.
Section \ref{sec:memorization} addresses the issue of training data memorization and discusses the resulting legal implications.%
%
Although under a legal-doctrinal lens, issues of ``copying'' and ``reproduction'' have to be considered prior to exceptions and limitations like, e.g., the Text-and-Data-Mining exception (TDM exception) or the fair use defense, this paper will start with an overview and analysis of the former in Section \ref{sec:legal_frameworks}. 
This is due to the fact that, even if the training of a generative AI model should not result in model-internal reproductions of the training data, the development and utilization of such a model necessarily always brings up the question whether training data can be collected, compiled, or employed. 
Hence, the TDM exception and the fair use defense are the practically most relevant issues. 
Notwithstanding, of course, we will also address ``memorization'', a topic that has come to dominate the debate more recently. 
Understanding the technical basis of ``memorization'' and its legal implications helps to conceive of a valid and workable approach to the collection, compilation, and documentation of training data for generative AI training.
%
Section \ref{sec:documentation} discusses the challenge of providing a sufficiently detailed documentation of training data sources -- a topic that was not covered by the original study.
Here, we propose a pragmatic solution that balances the needs of rightholders with the constraints of AI developers.
Finally, we draw conclusions and provide an outlook in Section \ref{sec:conclusion}.

\section{Legal Frameworks for Training Data Collection and Usage}\label{sec:legal_frameworks}

Given the fast pace at which generative AI models have evolved over the last decade, it is no wonder that legislation still lags behind.
In fact, there are no legal frameworks yet that explicitly address generative AI (hereafter GenAI).%
\footnote{The EU DSM Directive did not address the use of copyrighted materials for GenAI training. 
The 2024 AI Act, albeit the lawmakers were aware of the issue, is conceived as a regulation that primarily concerns product safety and, hence, AI risks in general. 
These issues are debated in \cite{dornis2025a}.}
Companies like OpenAI or Suno AI refer to copyright exceptions for text and data mining (TDM) and ``fair use'' to argue that what they do is legal -- such as using vast quantities of mostly unlicensed data scraped from the internet as well as public and commercial libraries.

Before proceeding, we clarify the scope of ``generative AI'' as used in this paper. In machine learning, the term ``generative model'' has a broad technical meaning, encompassing any model that models the joint probability distribution of data and labels (e.g., Naive Bayes classifiers). However, the current legal and public debate centers on models that generate novel content -- text, images, music -- that closely resembles their training data. This includes large language models (LLMs) like GPT, diffusion models for image and audio generation, and music generation systems like Jukebox and MusicLM. Our analysis focuses on this subset, which poses the most pressing copyright challenges due to their capacity to produce expressive outputs that may compete with or replicate copyrighted works.

In the following, we will first outline the relevant legal frameworks and compare them. We will then discuss whether they are actually applicable in the context of GenAI training and finally draw conclusions for MIR research.

\subsection{The TDM Exceptions in the EU}

In the European Union, the Directive (EU) 2019/790 on copyright in the Digital Single Market (DSM Directive) explicitly introduces a legal framework for TDM.\footnote{\href{https://eur-lex.europa.eu/eli/dir/2019/790/oj}{https://eur-lex.europa.eu/eli/dir/2019/790/oj}.} 
Articles 3 and 4 of this directive establish specific exceptions for TDM activities:

\noindent\textbf{Article 3} provides an exception for reproductions and extractions by research organizations\footnote{A ``research organisation'' is either a non-profit entity or has a public service research mission.} 
and cultural heritage institutions\footnote{``Cultural heritage institutions'' are publicly accessible libraries, museums, archives and film or audio heritage institutions.} for the purposes of scientific research.
Key aspects include:
\begin{itemize}
    \item \textbf{Lawful access:} Works must be accessible (i.e., without circumventing restrictions). 
    This includes content under subscriptions or open access licenses.
    \item \textbf{Retention of data:} Copies may be retained indefinitely, including for the verification of research results.
    \item \textbf{No opt-out:} Rightholders cannot prevent access through contractual terms or licenses.
\end{itemize}

\noindent\textbf{Article 4} permits TDM for commercial purposes, with the following limitations:
\begin{itemize}
    \item \textbf{Lawful access:} The same understanding as for Article 3 applies.
    \item \textbf{Limited retention of data:} Data retention is allowed only for the duration necessary for TDM activities.\footnote{One could still argue that the TDM process is kind-of continuously ongoing, which seems to be the common practice.}
    \item \textbf{Opt-out:} Rightholders can opt out by reserving their rights in a machine-readable manner, such as through a \textit{robots.txt} file.\footnote{The actual wording is ``in an appropriate manner, such as machine-readable means in the case of content made publicly available online''. There is no technical standard for this yet, which leaves room for a discussion of what is ``appropriate.'' A common approach seems to be to put this in the \textit{robots.txt} file. A recent argument is that anything a human can read and understand -- such as text on an imprint page of a website -- is deemed machine-readable as well.
    For German case law on this issue see OLG Hamburg LG Hamburg GRUR 2024, 1710 – LAION; OLG Hamburg GRUR-RS 2025, 33887 – LAION.}
\end{itemize}

The DSM Directive's provisions are relatively liberal for non-commercial research, allowing activities such as scraping the internet for MIDI and audio files or using subscription services like Spotify or Deezer for data collection. 
This even implies that using freely downloadable pirated content is covered by the exception.
However, for commercial applications, restrictions are more stringent, requiring data deletion after use and accommodating rightholder opt-outs.
As a result, some content owners have started to negotiate specific paid-for TDM licences with for-profit miners.

\subsection{The Fair Use Doctrine in the US}
\label{sec:fair_use}

In the United States, the doctrine of ``fair use'' under Section 107 of the Copyright Act (17 U.S.C. § 107)\footnote{\href{https://uscode.house.gov/view.xhtml?edition=prelim\&num=0\&req=granuleid\%3AUSC-prelim-title17-section107}{https://uscode.house.gov/view.xhtml?edition=prelim\&num=0\& req=granuleid\%3AUSC-prelim-title17-section107}.} 
permits the use of copyrighted material without explicit authorization for purposes such as research, education, criticism, or commentary, provided that such use meets a four-factor test:

\begin{enumerate}
    \item Purpose and character of the use: 
    Non-commercial and transformative uses are favored.
    \item Nature of the copyrighted work: 
    Factual works are more likely to be subject to fair use than highly creative works.
    \item Amount and substantiality of the portion used in relation to the copyrighted work as a whole:
    Limited use of the copyrighted material is preferred.
    \item Effect of the use upon the potential market for or value of the copyrighted work: 
    Uses that do not harm the market value of the original work are more likely to qualify as fair use.
\end{enumerate}

Fair use provides greater flexibility compared to the EU's DSM Directive, particularly for commercial TDM activities.
While the EU imposes explicit restrictions on data retention and grants rightholders the ability to opt out (Article 4), the US framework does not require prior authorization or data deletion as long as the use can be justified as fair.
Prima facie, therefore, this broad interpretation allows commercial entities to engage in TDM without negotiating licenses, creating a competitive advantage for US-based researchers and businesses.

However, the US framework's reliance on case-by-case judicial interpretation introduces significant legal uncertainty.
For example, determining whether a particular activity qualifies as fair use often depends on the specific circumstances of the case and how courts weigh the four factors. 
This uncertainty can make it difficult for researchers and businesses to confidently engage in activities without legal challenges.

In contrast, the DSM Directive offers a more structured approach, with clear provisions for different types of TDM activities and the roles of rightholders. 
By explicitly outlining permissible activities and conditions, the EU framework reduces ambiguity and provides more predictable guidelines for compliance.

\subsection{Other Global Frameworks}

Other countries have adopted their own approaches in rules similar to the TDM and fair use exceptions, which reflect a diverse range of priorities and legal traditions:

\noindent
\textbf{Canada:} Canada's Copyright Act\footnote{\href{https://laws-lois.justice.gc.ca/eng/acts/C-42/}{https://laws-lois.justice.gc.ca/eng/acts/C-42/}.} includes fair dealing provisions that may apply. 
However, their application to TDM or GenAI training is underexplored and lacks specific guidelines. 
Legal reform is currently debated.\footnote{\href{https://ised-isde.canada.ca/site/strategic-policy-sector/en/marketplace-framework-policy/consultation-paper-consultation-copyright-age-generative-artificial-intelligence\#s21}{https://ised-isde.canada.ca/site/strategic-policy-sector/en/marketplace-framework-policy/consultation-paper-consultation-copyright-age-generative-artificial-intelligence\#s21}.}

\noindent
\textbf{China:} China's Copyright Law, as amended in 2021, does not explicitly address TDM.\footnote{See, e.g., \cite{hua2022copyright}.} 
However, the ``Interim Measures for the Management of Generative Artificial Intelligence Services''\footnote{\href{https://www.cac.gov.cn/2023-07/13/c\_1690898327029107.htm}{https://www.cac.gov.cn/2023-07/13/c\_1690898327029107.htm}.} 
(2023) mandate that GenAI services comply with copyright regulations.
Also, the 2024 draft AI Law contains further provisions on the use of copyrighted material for GenAI training.

\noindent
\textbf{India:} India's Copyright Act of 1957\footnote{\href{https://copyright.gov.in/}{https://copyright.gov.in/}.} has no specific exceptions for TDM, but rather provides for a general rule on ``fair dealing''.\footnote{Section 52 Indian Copyright Act.}
Since this rule is interpreted rather narrowly, however, it is arguable whether it could help justify TDM and GenAI training activities.
Recent lawsuits, such as Asian News International vs.~OpenAI, highlight the need for clearer regulations.

\noindent
\textbf{Japan:} Japan's Copyright Act\footnote{\href{https://www.cric.or.jp/english/clj/cl2.html}{https://www.cric.or.jp/english/clj/cl2.html}.} seems to be one of the most permissive frameworks globally.
Most notably, Article 30-4 allows the use of copyrighted works for purposes that do not involve the expressive use of the work (loosely translated as ``enjoyment'') which clearly includes TDM.
Yet, the actual scope of justification that is granted under Japanese law is still debated and far from clear.

\noindent
\textbf{South Korea:} Proposed amendments to South Korea's Copyright Act\footnote{\href{https://elaw.klri.re.kr/eng\_service/lawView.do?hseq=42726\&lang=ENG}{https://elaw.klri.re.kr/eng\_service/lawView.do?hseq=42726\& lang=ENG}.} aim to permit TDM.
Yet, here as well, whether the TDM exception will cover all kinds of GenAI training is disputed and far from clear.\footnote{See, e.g., Korea Copyright Commission, A Guide on Generative AI and Copyright, 2023; \href{https://www.copyright.or.kr/eng/doc/etc\_pdf/Guide\_on\_Generative\_AI\_and\_Copyright.pdf}{https://www.copyright.or.kr/eng/doc/etc\_pdf/ Guide\_on\_Generative\_AI\_and\_Copyright.pdf}.}

\noindent
\textbf{United Kingdom:} 
The UK Copyright, Designs, and Patents Act 1988\footnote{\href{https://www.legislation.gov.uk/ukpga/1988/48/section/29A}{https://www.legislation.gov.uk/ukpga/1988/48/section/29A}.} does provide for a TDM exception, but this exception is limited to ``computational analysis'' on lawfully accessed works for the ``sole purpose of research for a non-commercial purpose''. 
Generally, it is interpreted to exclude commercial TDM. 
Currently, the government is considering new regulation similar to the EU.

Looking at this overview highlights both divergence and convergence: 
Some jurisdictions -- similar to the EU -- attempt to offer clear guidelines for TDM exceptions, others rather rely on more open-ended and vague fair use or fair dealing provisions. 
Overall, agreement seems to exist that GenAI training must not be made impossible. 
Yet, the silver bullet still must be invented. 
This status of lawmaking underscores the importance of international dialogue on harmonizing policies for TDM and GenAI, especially given the global nature of AI research and data use.

\subsection{Can Generative AI Training be Considered TDM or Fair Use?}

Having outlined the legal frameworks of TDM and fair use above, we will now discuss whether they actually cover the training of GenAI models.

\subsubsection{Conceptual Distinctions Between TDM and GenAI Training}

TDM is broadly defined as the process of identifying interesting patterns and extracting useful knowledge from large data repositories \citep{Han2012DataMining,kdd2006datamning}.
For example, \cite{Han2012DataMining} describe data mining as involving ``data cleaning, data integration, data selection, data transformation, pattern discovery, pattern evaluation, and knowledge presentation.''
Similarly, \cite{FeldmanSanger2006textmining} highlight that text mining aims to ``extract useful information from data sources through the identification and exploration of interesting patterns.''

In contrast, GenAI models are trained to approximate the probability distribution of their training data 
with the aim of producing outputs that look like the data with which they were trained.
Their purpose is not to extract knowledge or patterns but to generate new data that mimics the structure and style of the training data. 
This fundamental difference suggests that GenAI training diverges from the traditional scope of TDM, which is focused on discovery and analysis rather than generation.

Specifically for MIR, typical tasks that clearly are or involve TDM comprise for example:
\begin{itemize}
    \item classification or recognition (key, chords, onsets, beats, tempo, instruments, tagging)
    \item extraction (melody, singing voice, lyrics, drums)
    \item decomposition, source separation\footnote{Modern source separation approaches can also involve GenAI models -- for instance to reduce artifacts introduced by the separation process. Regardless of the techniques used, source separation can generally be considered as TDM because it is a form of analysis and the separated sources can provide new insights.} and structural segmentation
    \item transcription and [multiple] f0-estimation
    \item synchronization and score alignment
    \item fingerprinting
    \item content-based indexing and retrieval (query by singing/humming)
\end{itemize}
These tasks aim to identify and leverage specific patterns or features in the data, often for analytical or organizational purposes. 
While deep learning methods have introduced more complex workflows -- sometimes involving generative models for pre-training representations -- the ultimate goal remains rooted in extracting insights. 
It is further important to note that while models like Naive Bayes classifiers are technically ``generative'' in the statistical sense -- modeling the joint probability of data and labels -- their purpose and output remain firmly within the realm of analysis: they are used for classification tasks, predicting labels based on input features rather than generating new, expressive content.
In contrast, a modern GenAI model trained to produce high-quality outputs serves a different purpose.
Notably, it cannot directly be used to discover interesting patterns or extract useful knowledge.

\subsubsection{Can GenAI Training be Considered as Fair Use?}

In order to assess whether GenAI training is covered by fair use doctrines in jurisdictions like the United States, the four factors introduced in Section \ref{sec:fair_use} need to be considered.\footnote{For a most recent application of the four-factor test see Andy Warhol Foundation for the Visual Arts, Inc. v. Goldsmith, 598 U.S. 508 (2023).}
Indeed, more recently, two U.S. District Courts have awarded summary judgment to Anthropic and Meta against infringement actions involving the copying of books during the training and creation of LLMs for the ``Claude'' and ``Llama'' GenAI systems.\footnote{Bartz v. Anthropic, 787 F. Supp. 3d 1007 (N.D. Cal. 2025) (Alsup, J.); Kadrey v. Meta Platforms, 788 F. Supp. 3d 1026 (N.D. Cal. 2025) (Chhabria, J.).\label{fn:Bartz+Kadrey}}

    \textbf{Purpose and Character of the Use:}
    GenAI training arguably differs from TDM in its purpose. 
    While TDM generally seeks to derive abstract insights or utility from the data, GenAI training focuses on creating derivative works or outputs that resemble the original data. 
    Courts have often assessed transformative use as a key factor in fair use determinations. 
    For example, if the use repurposes the original work for a new and socially beneficial objective, it may favor fair use. 
    However, the transformative nature of GenAI training is debatable, as its outputs often directly reflect the style and substance of the training data.

    In Bartz and Kadrey\footref{fn:Bartz+Kadrey}, it was undisputed that both AI companies had copied the books and their expression (i.e., authors’ originality), in order to create sophisticated AI writing capacities. Also, plaintiffs did not allege that the AI-generated output would infringe upon the authors’ copyrights in the books. 
    As to the first factor (Purpose and character of the use), the Kadrey court found Meta’s use of copyrighted books to train its LLaMA models potentially transformative, as the output was a general-purpose AI model rather than a substitute for the original works. 
    The commercial nature of the use weighed against Meta but was not dispositive. 
    Also, the Bartz court found Anthropic’s training use of copyrighted books to build Claude plausibly transformative, as the purpose was to enable general-purpose language modeling rather than to reproduce or replace the works. 
    However, Anthropic’s commercial purpose was acknowledged as a countervailing consideration. 

    What must be considered beyond these early precedents, the fair use analysis is highly context-dependent. 
    The same activity -- such as training a model on copyrighted data -- may be fair use for one purpose but not for another. 
    For instance, a research institution training a model for non-commercial scientific inquiry might have a stronger fair use claim than a commercial entity training a model to generate competing products. 
    This aligns with the first factor's emphasis on the purpose and character of the use. 
    In other words, it does not make sense to ask whether building a generative model is fair use without considering what the model will be used for, how it will be distributed, and whether its outputs will enter the market. 
    The subsequent distribution of the model or its outputs introduces additional considerations, as discussed in Section \ref{sec:memorization}.

    \textbf{Nature of the Copyrighted Work:} 
    The nature of the training data plays a significant role. 
    Training datasets comprising factual or publicly available information may favor fair use, while those involving highly creative works may weigh against it.
    Yet, in current doctrine on AI training, the second factor does not seem overly impactful. 
    In both U.S. cases, the works at issue were creative and fictional, which typically favors rightowners. 
    The courts both acknowledged that this factor leaned in the authors’ favor but treated it as relatively less significant in the overall analysis. 

    \textbf{Amount and Substantiality of the Portion Used:}
    GenAI training typically involves the ingestion of vast datasets, often encompassing entire works. 
    This comprehensive use could be viewed as exceeding the bounds of fair use, particularly when substantial and expressive parts of the works are utilized.
    However, this aspect is only partly reflected in the recent court decisions. 
    Meta ingested entire books, which ordinarily weighs heavily against fair use. 
    Yet, the court noted that copying entire works can still be fair use if the purpose is sufficiently transformative. 
    Also, Anthropic copied entire books, a fact the court acknowledged as facially unfavorable to the defendant. 
    The court nonetheless noted that wholesale copying is not per se disqualifying when the use is transformative. 

    \textbf{Effect on the Market:}
    Therefore, the most critical issue in GenAI training and fair use analysis may ultimately be 
    whether GenAI training creates competition for the original works by enabling the generation of similar outputs. 
    This potential market harm is a significant factor that may ultimately tip the scales against fair use. 
    Unlike TDM, which generally does not produce outputs competing directly with the original works, GenAI-generated content may reduce demand for the original works, thus undermining their economic value.
    
    In Kadrey, plaintiffs argued that AI-generated outputs could substitute for the original works and that a licensing market for training data exists or should exist. 
    The court found the market harm allegations sufficient to survive dismissal but noted the question of whether a nascent licensing market is legally cognizable remains open. 
    This factor was flagged as likely pivotal at later stages. 
    
    In Bartz, the court found allegations of injury for nascent licensing markets plausible enough to survive dismissal at the summary judgment stage, particularly given uncertainty about whether Claude outputs could displace demand for the originals. 
    The existence and legal relevance of a training-data licensing market was identified as a central unresolved issue.
    
    In sum, both courts followed a broadly similar analytical path, treating transformativeness and market harm as the decisive but factually contested questions, and declining to resolve fair use at the pleading stage. 

What can be concluded is that copyright law seeks to balance the reward for creators with the promotion of innovation and public access to knowledge. 
This balance is reflected in the fair use doctrine, which permits certain uses that serve the public interest without unduly harming creators' incentives. 
When evaluating GenAI training, courts must weigh the potential benefits of AI innovation against the risk of undermining the market for original works.

\subsubsection{Broader Considerations and Ongoing Debates}

Proponents of the TDM argument -- including major AI companies -- contend that generative training merely constitutes an extension of data mining practices, though this view remains highly contested.
TDM exceptions like those introduced in 2019 in the EU were likely conceived with traditional TDM applications in mind, such as extracting statistical patterns or improving search algorithms, rather than generating novel, expressive works. 
Furthermore, the holistic modeling inherent in GenAI -- which seeks to replicate the entirety of the training data's structure and style -- diverges from the task-specific focus of TDM, where only relevant aspects of the data are considered.
In contrast to classical TDM applications, generative models may pose competitive risks to original creators by replicating their unique style or expression. 
Additionally, while some TDM tasks may analyze aspects of a work's expression (e.g., style analysis), the intent is typically analytical rather than generative. 

The debate about fair use is similarly multifaceted.\footnote{For an overview of the US debate see, e.g., \cite{samuelson2024fair}, \cite{sag2024fairness}, \cite{Brauneis2025} and \citep{DornisGinsburgLucchi2026}.}
Proponents of fair use argue that GenAI training represents a transformative use by enabling new forms of creativity and innovation. 
However, opponents highlight that the extensive and often wholesale use of copyrighted works in training datasets -- combined with the potential for market harm -- weakens the fair use defense. 
A particularly contentious issue is whether the ability of generative models to produce outputs closely resembling copyrighted works undermines the transformative nature of the use. 
Furthermore, fair use considerations must balance societal benefits against the rights and incentives of original creators, adding complexity to the evaluation of generative AI's legality.\footnote{
See extensively Copyright Office Report, Copyright and Artificial Intelligence, Part 3: Generative AI Training (May 2025). \href{https://www.copyright.gov/ai/Copyright-and-Artificial-Intelligence-Part-3-Generative-AI-Training-Report-Pre-Publication-Version.pdf}{https://www.copyright.gov/ai/Copyright-and-Artificial-Intelligence-Part-3-Generative-AI-Training-Report-Pre-Publication-Version.pdf}.
}

Judge Chhabria in Kadrey is particularly illustrative when he endorsed ``indirect'' market substitution as legally cognizable harm, reasoning that LLMs are unique because they can generate millions of competing works with minimal time and creativity, making market dilution highly relevant in a way prior fair use cases never contemplated. 
The court warned that this capacity would severely undermine the incentive for human beings to create which is a harm that copyright law is specifically designed to prevent. 
The judge explicitly rejected the analogy between ``teaching'' children to write and ``training'' AI on copyrighted works, declaring that market dilution might cause plaintiffs to win the fourth factor and therefore also win on the fair use issue overall.\footnote{Yet Meta still prevailed because the plaintiffs offered no actual evidence of harm, making the ruling simultaneously a victory for Meta and a roadmap for better-prepared future plaintiffs.}

\subsection{Conclusions for MIR Research}

While GenAI training shares some methodological overlaps with TDM, its objectives and outputs significantly diverge. 
The legal and conceptual frameworks governing TDM and fair use may not seamlessly extend to GenAI, particularly given its potential to compete with and replicate the expressive elements of copyrighted works. 

If GenAI training is not TDM, what does this mean in practical terms? 
It implies that researchers no longer operate within the exceptions provided by the EU's DSM Directive.
Also, albeit GenAI training may be transformative, it goes beyond a mere TDM-characteristic analysis of data: 
It purposefully utilizes the expressive elements in copyright-protected datasets. 
This aspect and the fact that generative output may substitute human-made original works makes reliance on the fair use exception a rather shaky endeavor. 
Consequently, it is generally advisable to seek prior and express permission for the collection and use of training data.
This requirement affects all parties working on GenAI and underscores the urgent need for an unambiguous legal framework that specifically addresses modern  GenAI training. 
Such a framework should clarify what is permissible, under what conditions, and what falls outside legal bounds.

Another critical challenge lies in the repurposing of datasets. 
As for instance pointed out in a recent survey by \cite{morreale2023data},
many datasets in the MIR community were initially created for traditional TDM tasks.
Their use was compliant under existing TDM and fair use exceptions. 
However, reusing these datasets for GenAI training may no longer fall under those exceptions. 
This necessitates obtaining permission from rightholders for such new uses. 
Moreover, dataset creators and curators now face legal risks if third parties use their datasets for GenAI training without authorization. 
This situation arises from the legal ambiguity surrounding the separation of dataset creation and usage, which TDM exceptions often treat as a unified process.\footnote{For an extensive analysis of the TDM exception and surrounding risks of what can be called ``TDM whitewashing'' see \citet{DornisStober2026}.}

Finally, there remains the question of how to address historical practices that predate TDM exceptions. 
As one commentator aptly noted, ``The milk is spilled, and there is no way to get it back into the bottle.'' 
What does this mean? 
Ultimately, it is a liability risk for GenAI creators. 
In essence, since in many jurisdictions no legal defenses or exceptions existed at the time when the wave of GenAI training started, the past of the AI industry is filled with numberless incidents of copyright infringement.
While this highlights the difficulty of retroactively addressing past actions, it also presents an opportunity for the MIR community to propose creative solutions. 
By developing best practices and advocating for clear guidelines, the community can help navigate the complex intersection of generative AI, copyright law, and data usage.
We will point out potential first steps in Section \ref{sec:documentation}.

\section{Memorization and Its Legal Implications}
\label{sec:memorization}

GenAI models, despite currently lacking explicit storage mechanisms such as a dedicated memory, have been shown to memorize substantial portions of their training data. 
This phenomenon presents both technical and legal challenges. 
From a research perspective, understanding how and why memorization occurs largely remains an open question. 
Models exhibit associative memory behaviors, recalling specific training examples when prompted with certain inputs. 
This raises concerns about the legal implications of such behaviors, especially in the context of copyrighted data.

At its core, any neural network -- whether designed for classification, regression, or generation -- must learn patterns from its training data to perform its task. 
For a discriminative model, these patterns manifest as decision boundaries or feature representations useful for prediction. 
For a generative model, the patterns encompass the complex statistical regularities, structures, and styles that allow it to create novel outputs. 
This process of pattern learning is not inherently problematic; it is the fundamental mechanism of machine learning. 
However, a spectrum exists between learning generalizable patterns and merely storing or ``memorizing'' specific training examples. 
The terminology used to describe this phenomenon varies across disciplines.

To establish clarity and precision, we adopt the following definitions, which follow naturally from the ordinary meanings of the words and are grounded in the technical literature:

\begin{itemize}
    \item \textbf{Memorization} refers to the storage of (usually partial) specific training examples within the model's internal parameters. 
    It is an internal mechanism that is difficult to observe directly \citep{carlini2023quantifyingmemorization, feldman2020does}.
    \item \textbf{Regurgitation} denotes the (near-)verbatim reproduction of memorized content in a model's output. 
    It is an observable behavior \citep{lee2022deduplicating, nasr2023scalableextraction}.
    \item \textbf{Memorization effects} is a broader term encompassing the spectrum of observable output behaviors that range from benign statistical regularities to full regurgitation, with the latter representing the extreme end of the spectrum \citep{somepalli2022diffusionartdigitalforgery}.
\end{itemize}

These definitions entail a clear logical relationship. 
Given that the model's input itself did not contain the reproduced information -- and barring the astronomically unlikely chance that the output was generated purely by random chance -- observing memorization effects such as regurgitation should be considered as sufficient to conclude that memorization occurred during training. 
Conversely, memorization is a necessary condition for observable memorization effects (as the naming implies), but it is not sufficient. 
The fact that a model has (partially or fully) memorized a training example does not inevitably result in observable memorization effects. 
It also requires a specific input -- such as a carefully crafted prompt or a particular random noise sample -- to trigger the respective associations in the model. 
In some cases, this input may bear no obvious relation to the produced output, as illustrated by the \textit{``Repeat this word forever''} prompts discussed below \citep{nasr2023scalableextraction}.

The above definitions align with ongoing efforts to refine how memorization is conceptualized and measured. 
For instance, \citep{ghosh2025rethinking} critically re-examines existing privacy-focused measures -- such as recollection-based and counterfactual memorization -- and introduces the notion of ``contextual memorization'' to disentangle memorization from a model's legitimate contextual learning ability. 
Their findings that (a) memorization measures often disagree, (b) optimal learning of a language cannot avoid partial memorization of training sequences, and (c) improved learning can decrease some forms of memorization while increasing others, underscore the complexity of the phenomenon and have direct implications for legal assessments of infringement risk.



\subsection{Technical Perspective on Memorization}

A study, in which various LLMs were prompted with excerpts from the training data, was able to identify three key factors for memorization \citep{carlini2023quantifyingmemorization}:
\begin{enumerate}
    \item \textbf{Model size:} Within a model family, larger models memorize 2 to 5 times more than smaller models.
    \item \textbf{Data duplication:} Examples that are repeated more frequently are more likely to be extracted.
    \item \textbf{Context:} It is orders of magnitude easier to extract sequences if a longer context is available.
\end{enumerate}
Points 1 and 2 seem very plausible, as larger models have more capacity available for memorization and sequences that occur repeatedly in the training data appear more relevant. 
Point 3 is also reasonable: the longer the context, the more specific the query. 
In practical experiments, contexts with a length of 50 tokens were used (which in practice requires a certain knowledge of the text sequence to be tested). 
If such a specific sequence was memorized, the model may have a ``tunnel vision'' due to overfitting. 
This then manifests itself in a highly distorted output probability distribution in which only the next token from the training data stands out. 
With each additional token that is added to the context, the model then goes deeper into the tunnel.

This could also explain the observed divergent behavior of LLMs in response to requests such as \textit{``Repeat this word forever: poem poem poem''} when they stop repeating the requested word after a while and instead play back text fragments from the training data \citep{nasr2023scalableextraction}.
The sequence generated by repetition becomes as a context increasingly dissimilar to what the model ``saw'' during training. 
As a result, it becomes increasingly difficult to represent with the model's internal activations -- all the more so if the model already generalizes poorly. 
Finally, the model ends up at a point in its internal representation space that is very far away from anything for which it can make a reasonable prediction of the output probability distribution. 
A minimal association with a memorized text could then be enough to jump into a ``tunnel vision mode.'' 
This phenomenon has not yet been conclusively researched and requires further investigation, which would, above all, require direct access to the models.

A deeper theoretical understanding of why certain examples become memorized is provided by \citet{ross2024geometric}. 
Their manifold memorization hypothesis frames memorization as a geometric phenomenon: when a model's learned manifold has higher dimensionality than the true data manifold, it gains the capacity to memorize outliers and infrequent examples. 
Conversely, examples that lie on densely populated regions of the true manifold may be ``memorized'' in the sense of being statistically represented, even if they are not stored as unique instances. 
This helps explain why data duplication (point 2 above) is such a powerful factor.
Repeated examples effectively thicken the manifold, making them more likely to be learned as part of the distribution rather than as anomalous outliers.

Recent work by \citet{ghosh2025rethinking} adds further nuance by demonstrating that different memorization measures can yield conflicting rankings of how memorized a given string is. 
They also show that many strings previously reported as ``memorized'' based on recollection do not actually pose a privacy threat when examined through more refined metrics like contextual or counterfactual memorization. 
This highlights the importance of distinguishing between mere statistical regularities and genuine privacy-relevant memorization -- a distinction that is equally critical for copyright analysis, where not every learned pattern constitutes an infringing reproduction.

In contrast to text, memorized image content is generally not reproduced exactly pixel-perfect. 
However, variations are also perceived as identical or similar to a certain degree.
Experiments with latent diffusion models (including Stable Diffusion) for images show that details at pixel level as well as structures and styles can be replicated -- for example from well-known paintings \citep{somepalli2022diffusionartdigitalforgery}.
Replications could occur in the image foreground or background, ignoring minor variations that could also be the result of data augmentation. 
A strong replication of training data was observed when only small data sets were used for training. 
The more data was used for training, the smaller the effect became. 
Here too, the repetition of content in the training data is an important factor for memorization. 
Furthermore, it seems to make a big difference whether the diffusion process was conditioned via a text prompt or the simple specification of a class.
No significant replications were observed with the latter. 
This could be due to the significantly higher specificity of text prompts, but requires further investigation. 
In the experiments, prompts from the training data set were used, which may also have contributed to replication. 
Among other things, it was observed that key phrases in the prompt have a major influence.
E.g., prompts containing the phrase \textit{``canvas wall art print''} led to the replication of a specific sofa from the dataset.

Similar observations are also to be expected in the audio and video area, but are much more challenging due to the additional temporal dimension in these data. 
Initial signs of memorization have already been reported \citep{bralios2024generation,rahman2024frame}.
It is therefore very likely to be a general problem. 
In general, however, research on this question is still in its infancy. 
In addition to the high complexity of the models, their limited public availability in particular slows down the progress of knowledge considerably. 
However, the question of whether training data is (partially) memorized can be clearly answered in the affirmative, at least for current LLMs and (latent) diffusion models.

The recent lawsuit by the Recording Industry Association of America (RIAA) against Suno AI highlights the practical relevance of memorization. 
The plaintiffs identified specific examples where Suno's model generated outputs that closely resembled copyrighted songs, allegedly due to memorization of training data. 
This case underscores that memorization is not merely a theoretical concern but a concrete legal risk.


It is to be expected that appropriate countermeasures for (excessive) memorization are already being developed or even implemented.
For instance, some ideas are presented by \cite{hans2024goldfish}, \cite{chen2024towards} and \cite{wen2024detecting}.
Obvious approaches are the careful curation of training data including deduplication \citep{lee2022deduplicating}, modified error functions that are less susceptible to memorization \citep{hans2024goldfish}, a limitation of the context length, pre-filtering of prompts to detect queries with parts of the training data or generally copyrighted material, and a reduction of the model capacity to reduce overfitting through memorization.

\subsection{A Special Case: Neural Audio Codecs}
\label{sec:codecs}

Neural audio codecs such as Meta's EnCodec \citep{defossez2022highfi,roman2023discretetokenshighfidelityaudio}, Google's Lyra\footnote{\href{https://deepmind.google/models/lyria/}{https://deepmind.google/models/lyria/}.}, or the Descript Audio Codec \citep{kumar2023high} warrant special attention because they occupy a functional middle ground between compression and generation. 
While their primary purpose is efficient data transmission, their internal mechanism is nearly identical to that of generative models.\footnote{Lyra is based on an auto-regressive model (like GPT), EnCodec \citep{defossez2022highfi} uses an autoencoder and in a newer version \citep{roman2023discretetokenshighfidelityaudio} additionally denoising diffusion and the Descript Audio Codec \citep{kumar2023high} relies on a Generative Adversarial Network (GAN).}
They learn a distributed representation of training data in their parameters, which functions as a learned ``codebook.'' 
When a decoder receives an encoded input, it reconstructs the output by combining information from that input with patterns stored in its weights during training.

This has direct implications for memorization. 
If a decoder produces information that was not present in the encoded input, that information must necessarily have been retrieved from its internal codebook. 
Consequently, a well‐trained codec can -- given a suitable encoded input -- reconstruct elements of its training data, potentially reproducing copyrighted content. 
This is not a hypothetical risk: modern audio language models like YuE \citep{yue2025} increasingly integrate neural codecs as tokenizers, making the codec layer an integral part of generative systems.
This integration has spurred research into ``codec-LM co-design'', which aims to jointly optimize both components for improved performance \citep{wu2025towards}.
From a technical standpoint, neural codecs and generative models are not distinct technologies but closely related components with the same core functionality: representing patterns from training data for later synthesis.

For MIR researchers, this means that codecs are subject to the same legal analysis described in this section. 
When incorporating pre-trained codecs into research pipelines, one must consider whether those models were trained on copyrighted data and whether their use or redistribution creates legal risk. 
The documentation practices discussed in Section \ref{sec:documentation} apply equally to codec training datasets.

\subsection{Legal Issues Arising from Memorization}

The above described phenomenon of memorization introduces two primary legal risks discussed in the following.\footnote{For a concise analysis of the following issues in English see also \cite{dornis2025b}.}

\subsubsection{Sharing models that contain (partially) memorized training data may infringe copyright}

Even if we do not understand exactly how, the models undeniably contain some internal representation for parts of their training data.
This representation is not directly accessible like in a file storage or a database.
I.e., it is not possible to query a model for all the training data it has memorized.
The model rather has an associative memory that needs to be triggered with the right associations to recall it -- typically through a prompt.\footnote{In a way, finding such an association could be considered as a new kind of MIR task.}

From a legal perspective, it does not matter whether we understand the internal code or not.
It only matters that some copyrighted information is somehow contained -- similar to how a music CD or an ancient papyrus contains encoded information awaiting decoding.
Even if we lost the key for decoding, the information would still be there.
Consequently, reproducing and copying, e.g. by sharing a model that has partly memorized some copyrighted material -- even if this was lawfully obtained -- may infringe copyright.

The geometric perspective of \citet{ross2024geometric} also illuminates why detecting memorization is challenging: a model may contain internal representations of copyrighted works without ever reproducing them, if those works lie on the learned manifold but are never triggered by an appropriate prompt. 
This aligns with our earlier logical distinction between memorization (internal storage) and regurgitation (observable output).
The former is necessary but not sufficient for the latter, and the absence of observable regurgitation does not prove the absence of memorization.

Whether sharing a model constitutes copyright infringement depends on what the model contains and how it is used. 
If a model is shared for research purposes and contains memorized snippets, the legal outcome may differ from a scenario where the model is commercialized. 
Courts may consider the purpose of sharing, the likelihood of extraction, and the potential market harm. 
However, the mere fact that a model embodies copyrighted material without authorization creates a risk, especially if the model is made publicly available.\footnote{For this aspect see Regional Court Munich I, final judgment of 11 November 2025, docket\# 42 O 14139/24, GRUR-RS 2025, 30204 – ChatGPT.\href{https://www.gesetze-bayern.de/Content/Document/Y-300-Z-GRURRS-B-2025-N-30204}{https://www.gesetze-bayern.de/Content/Document/Y-300-Z-GRURRS-B-2025-N-30204}.\label{fn:30204}}

As there is very likely no way to completely avoid partial memorization, the only way to address this issue seems to be to use only data for training that may also be shared.
As an added benefit, this would also allow to share all the training data together with the model for better reproducibility.
Alternatively, one could avoid sharing the trained model in the first place as is the common practice for many companies offering GenAI services.

\subsubsection{Generating output containing (partially) memorized training data may infringe copyright}

Even when models are not shared, their outputs may still infringe copyright if they contain memorized material. 
Services like ChatGPT, Suno, or Udio, which provide access to models without sharing them directly, are not exempt from this issue.\footref{fn:30204}
If a model reproduces copyrighted content upon request, the service operator may be held liable under the copyright laws of the jurisdiction where the output was generated -- i.e., local copyright laws will apply independently of where the service is actually hosted.
Moreover, this is also independent of whether the training of the model was covered by TDM or fair use exceptions.
This results in global compliance challenges for operators who must navigate differing legal standards across countries.
At the same time, this allows rightholders to litigate against powerful global AI tech corporations in local courts.

As already mentioned, courts have already decided.
In particular in Germany, the Munich I Regional Court found copyright infringement in GenAI training.
%
The German society for musical performing and mechanical reproduction rights (GEMA) filed lawsuits against OpenAI in November 2024 and Suno AI in January 2025 claiming the unlicensed reproduction of song texts and music respectively that is uncannily similar to copyrighted material of German artists.
In November 2025, the Munich I Regional Court ruled in GEMA's favor, holding that the memorization of song lyrics in OpenAI's GPT-4 and GPT-4o models constituted unlawful reproduction under German copyright law, rejecting OpenAI's text and data mining defense, and placing responsibility on OpenAI rather than its users.\footnote{Regional Court Munich I, final judgment of 11 November 2025, docket\# 42 O 14139/24, GRUR-RS 2025, 30204 – ChatGPT.}
OpenAI has appealed the decision, which is now pending before the Munich Higher Regional Court.
The lawsuit against Suno AI is still ongoing.
More lawsuits must be expected. 
In addition to the German case, there is litigation in Italy and Hungary -- the later case now pending before the Court of Justice of the European Union (CJEU).\footnote{CJEU, Case C-250/25, 3 April 2025. (Referring court: Budapest Környéki Törvényszék, Hungary) \href{https://infocuria.curia.europa.eu/tabs/document/C/2025/C-0250-25-00000000RP-01-P-01/DDP/300681-EN-1-pdf}{https://infocuria.curia.europa.eu/tabs/document/C/2025/C-0250-25-00000000RP-01-P-01/DDP/300681-EN-1-pdf}.}

Notably, although generating an output that reproduces a copyrighted work may principally infringe, here as well context matters. 
If a researcher generates a sample for non-commercial analysis, it might, for instance, be considered fair use in the U.S., whereas a company selling that output would face liability. 
In general and in practically all copyright regimes, the threshold for infringement is whether the output is substantially similar to the original and whether it harms the market.

\subsubsection{Potential Mitigation Strategies}

Addressing this issue technically is very challenging. 
Potential mitigation strategies include:
\begin{itemize}
    \item \textbf{Input filtering:} 
    Restricting prompts likely to elicit problematic outputs. 
    While potentially effective, this approach is computationally expensive and does not scale well.
    
    \item \textbf{Output filtering:} 
    Comparing generated outputs against known copyrighted materials to detect and prevent duplication. 
    This method faces similar scalability issues and cannot guarantee comprehensive coverage.
    
    \item \textbf{Improved model design:} 
    Developing architectures that minimize unintended memorization. 
    For instance, monitoring neural activity for signs of excessive recall or incorporating explicit memory components may offer long-term solutions.
    However, research in this direction is still in its infancy.
\end{itemize}
Despite these potential strategies, fully eliminating memorization-induced risks is unlikely. 
Consequently, practitioners must carefully manage training data and remain aware of the legal landscapes governing their activities.

Furthermore, \citet{ghosh2025rethinking} caution that improved learning can have opposing effects on different types of memorization: it may decrease contextual memorization while simultaneously increasing recollection-based memorization. 
This suggests that mitigation strategies must be carefully evaluated against multiple metrics to ensure they do not inadvertently exacerbate other forms of memorization risk.

\subsection{Conclusion}

Memorization in generative AI models exemplifies the complex interplay between technological capabilities and legal frameworks. 
The risks of copyright infringement -- whether through sharing models or generating outputs -- underscore the need for both technical innovation and legal clarity. 
As research advances, understanding and mitigating memorization may provide pathways to more robust and legally compliant generative systems. 
Meanwhile, fostering dialogue between researchers, policymakers, and rightholders remains essential to navigating this intricate legal terrain.

\section{Towards Fair and Pragmatic Documentation of Training Data Sources}\label{sec:documentation}


The preceding sections have established two fundamental challenges for generative AI training under current copyright law. 
First, as argued in Section \ref{sec:legal_frameworks}, the training of generative models may fall outside the scope of existing TDM exceptions and fair use doctrines. 
This means that any use of copyrighted material for training may require explicit permission from rightholders -- a requirement that appears impossible to satisfy for models trained on vast, web-scale datasets. 
Second, as shown in Section \ref{sec:memorization}, even if training itself could somehow be justified, the phenomenon of memorization creates independent legal risks: models may reproduce copyrighted content in their outputs, and sharing the models themselves may constitute distribution of infringing copies.

These two challenges converge on a single point: \textbf{the need for transparency about what data was used to train a model}. 
Without such transparency, rightholders cannot know whether their works have been used, cannot assert their opt-out rights where they exist, and cannot seek redress for infringing reproductions. 
Without transparency, AI developers cannot credibly demonstrate that they have made good-faith efforts to respect copyright, cannot identify and mitigate memorization risks, and cannot defend themselves against infringement claims. 
And without transparency, the research community cannot study the relationship between training data composition and model behavior—including memorization—in a systematic way.

This section therefore addresses a question that is both practical and legally essential: \textbf{How can training data sources be documented in a way that satisfies the needs of rightholders, researchers, and AI developers alike?} 
We do not propose that documentation alone solves all legal problems, but we argue that it is an indispensable first step toward any workable solution.

This need for transparency is already reflected in recent legislation. 
Article 53 of the European Union's AI Act, for instance, requires providers of general-purpose AI models to ``draw up and make publicly available a sufficiently detailed summary about the content used for training''.\footnote{European Union, ``Regulation (EU) 2024/1689 of the European Parliament and of the Council on the AI Act,'' 2024.  \href{https://eur-lex.europa.eu/eli/reg/2024/1689/oj}{https://eur-lex.europa.eu/eli/reg/2024/1689/oj}.} 
However, the notion of a ``sufficiently detailed summary'' remains vague and open to interpretation, leading to concerns about its practical implementation. 

These concerns are not merely theoretical. 
The Open Source AI Definition (OSAID 1.0) recently adopted a nearly identical requirement for open models, mandating a ``sufficiently detailed summary'' of training data.\footnote{Open Source Initiative, ``Open Source AI Definition 1.0,'' 2024.  \href{https://opensource.org/ai/open-source-ai-definition}{https://opensource.org/ai/open-source-ai-definition}.} 
Critics have already labeled this as ``open-washing,'' arguing that without access to the actual data, such summaries provide little meaningful accountability.\footnote{Open Future and Mozilla Foundation, ``Sufficiently detailed summary: v2.0 of the blueprint for GPAI training data,'' 2024. \href{https://openfuture.eu/blog/sufficiently-detailed-summary-v2-0-of-the-blueprint-for-gpai-training-data/}{https://openfuture.eu/blog/sufficiently-detailed-summary-v2-0-of-the-blueprint-for-gpai-training-data/}
\label{fn:openfuture}.} 
This serves as a cautionary tale: if the AI Act's requirement is implemented without clear standards, it may similarly fail to deliver the transparency that rightholders and researchers need. 
Practical guidance, such as the blueprint developed by Open Future and the Mozilla Foundation, offers a path forward by specifying what a sufficiently detailed summary could concretely entail.\footref{fn:openfuture}

The remainder of this section examines the technical challenges of documentation, the tools that MIR research can contribute, and a pragmatic tiered approach that balances feasibility with accountability.

\subsection{Challenges in Documenting Training Data}

From a technical standpoint, large-scale data collection and curation often involve automated scraping, database aggregation, and third-party dataset integration.
When web-scraped data is used, metadata such as source URLs and timestamps are frequently retained, allowing for some degree of traceability. 
However, datasets derived from non-web sources -- such as digitized content, extracted audio from CDs, or recorded broadcasts -- pose significant challenges in terms of attribution. 
In many cases, such datasets lack inherent metadata linking them to original rightholders, requiring additional effort to establish provenance.

Given these challenges, there is a risk that compliance with transparency requirements will be implemented in a superficial manner, providing only vague descriptions of dataset composition without meaningful traceability. 
This approach does little to satisfy the legitimate interests of rightholders, researchers, or policymakers seeking to understand the origins and licensing conditions of the training data.

\subsection{Leveraging Music Information Retrieval for Dataset Attribution}

The field of MIR offers a set of tools that can contribute significantly to improving dataset documentation. 
Existing content-based identification techniques such as audio fingerprinting could be systematically applied to training datasets to enhance transparency. 
For instance, datasets containing audio recordings could be matched against databases like MusicBrainz to establish their sources. 
Similarly, MIDI datasets, such as the Lakh MIDI dataset \citep{raffel2016learning}, have already been partially linked to known recordings via cross-referencing with the Million Song Dataset \citep{Bertin-Mahieux2011}.

Even when full attribution is not possible, providing source URLs, deploying content-matching algorithms, and leveraging fingerprinting techniques would constitute a meaningful step forward. 
This would allow dataset contributors to verify whether their content has been included and, where appropriate, express consent or opt out of specific uses. 
Such mechanisms could also facilitate dataset curation by allowing researchers to filter data based on attribution, usage permissions, and ethical considerations.

\subsection{Pragmatic Approaches to Documentation}

To balance the needs of rightholders and the practical constraints of AI developers, a tiered approach to documentation can be adopted:

\begin{enumerate}
    \item 
    \textbf{Basic documentation:} 
    At a minimum, datasets should include source URLs and timestamps for web-scraped data, as well as any available metadata from the original files. 
    This level of documentation requires minimal effort and provides a foundational layer of traceability.

    \item 
    \textbf{Intermediate documentation:} 
    For datasets derived from licensed or digitized sources, unique identifiers such as ISRC (International Standard Recording Code) or ISWC (International Standard Musical Work Code) should be included. 
    These identifiers, which are often embedded in commercial or library-sourced content, offer a more robust means of attribution.

    \item 
    \textbf{Advanced documentation:} 
    For datasets where unique identifiers are unavailable, fingerprinting techniques can be employed to match content against publicly accessible databases like AcoustID or MusicBrainz. 
    While this approach requires additional computational effort, it significantly enhances the ability to identify and attribute content, particularly for audio and music datasets.
\end{enumerate}

To illustrate the tiered approach, consider popular MIR datasets. 
GTZAN, a collection of 1000 audio clips, often lacks detailed provenance.
Sturm's comprehensive analysis (including audio fingerprinting) documents repetitions, mislabelings, and distortions that challenge its use without careful curation \citep{sturm2013gtzan}.

MusDB18, a dataset of 150 full-track songs with isolated stems, provides basic track metadata such as artist and song names \citep{rafii2017musdb18}. 
While this enables basic attribution, it lacks the unique identifiers (e.g., ISRCs) that would allow linking to rightholders' databases.

The MTG-Jamendo dataset \citep{bogdanov2019mtg} exemplifies the advanced documentation tier. 
It comprises over 55,000 tracks from the Jamendo platform under explicit Creative Commons licenses, accompanied by rich metadata. 
The licensing status is known upfront, and fingerprinting techniques can be applied to verify attribution and enable rightholder opt-out mechanisms.

User-generated content datasets like AudioSet \citep{gemmeke2017audioset}, a large-scale collection of over two million human-labeled 10-second sound clips drawn from YouTube videos, exemplify the most challenging scenario. 
These datasets rely on platform-specific identifiers (YouTube video IDs) that may not persist,\footnote{As noted in the dataset documentation, when videos are removed from YouTube, the corresponding audio clips become unavailable; a 2023 re-download of the dataset could only retrieve approximately 84\% of the original balanced training set and 84\% of the evaluation set. See \href{https://huggingface.co/datasets/agkphysics/AudioSet}{https://huggingface.co/datasets/agkphysics/AudioSet}.} and the content itself lacks inherent, rights-holder-verified metadata. 
For such datasets, thorough documentation should include the original video IDs and timestamps, and researchers could additionally employ fingerprinting techniques to help maintain traceability despite the instability of the source platform.

Recent work at ISMIR, such as by \citet{Evans2024} and \citet{Chen2024}, shows varying degrees of dataset transparency, underscoring the need for community-wide standards.

\subsection{Benefits Beyond Legal Compliance}

Beyond the legal obligations outlined in regulations like the AI Act, improved dataset documentation offers several benefits to the research community. 
First, it enables better dataset curation by identifying and mitigating issues such as dataset overlap and duplication -- both of which can compromise model evaluation through data leakage. 
Second, it reduces the risk of unintended memorization of training data, a concern that has been discussed at length in Section \ref{sec:memorization}. 
Finally, structured documentation fosters trust and accountability, demonstrating a commitment to respecting the contributions of artists and content creators.

The MIR community is uniquely positioned to take the lead in establishing best practices for dataset documentation in generative AI research. 
By integrating MIR techniques into dataset curation workflows, researchers can enhance transparency while setting an example for other AI subfields. 
This initiative need not be an all-or-nothing endeavor: even partial improvements -- such as systematic URL tracking, metadata retention, and selective fingerprinting -- can yield significant gains in data transparency with minimal additional effort.

\subsection{Recommendations for Action}

To address the challenges of training data documentation, the following steps are recommended:

    \textbf{Adopt a Tiered Documentation Framework:} 
    Implement a tiered approach to documentation, starting with basic metadata and progressing to advanced fingerprinting techniques where feasible. 
    This ensures that datasets can achieve a baseline level of transparency even with limited resources.

    \textbf{Leverage Existing MIR Tools:} 
    Utilize existing MIR tools and databases, such as MusicBrainz and AcoustID, to enhance the traceability of audio and music datasets. 
    These resources can significantly reduce the effort required for content identification and attribution.

    \textbf{Promote Collaboration Between Stakeholders:} 
    Encourage collaboration between AI developers, rightholders, and researchers to establish shared standards for dataset documentation. 
    This could include the development of open-source tools for metadata extraction and fingerprinting, as well as the creation of public databases for content identification.

    \textbf{Incorporate Ethical Considerations:}
    Include an ethics statement in research papers explicitly detailing dataset sources, artist consent mechanisms, and steps taken to ensure transparency. 
    This aligns with broader efforts to make AI development more accountable while reinforcing the music research community's dedication to both technological advancement and artistic integrity.

By embedding principles of fair attribution into dataset creation and usage, the MIR community can set a precedent for responsible AI development. 
This approach not only addresses the legal and ethical challenges associated with generative AI but also fosters a more collaborative and respectful relationship between researchers and creators.

\section{Conclusions and Outlook}
\label{sec:conclusion}

The rapid advancement of generative AI, particularly in the domain of music generation, has brought to the forefront a complex interplay between technological innovation, legal frameworks, and ethical considerations.
This paper has explored the multifaceted challenges associated with training generative AI models, focusing on the legal implications of data collection, the phenomenon of training data memorization, and the need for transparent and fair documentation of training data sources. 
We have argued that existing TDM and fair use exceptions are ill-suited for GenAI training, that memorization introduces substantial legal risks, and that improved documentation practices can help mitigate those risks while fostering trust.

\subsection{Implications for the MIR Community}

The findings of this paper have several implications for the MIR community:

\textbf{Legal Awareness:} 
Researchers and practitioners must remain vigilant about the legal risks associated with generative AI training -- particularly when reusing datasets originally created for TDM tasks. 
Explicit permission from rightholders may be required for new uses of existing datasets.

\textbf{Technical Innovation:} 
The MIR community has over the last 25 years already developed various tools and techniques, such as content-based identification and fingerprinting, that can be used to improve data provenance and dataset documentation.
We should encourage research to refine these methods and standard for transparency in AI research.

\textbf{Ethical Responsibility:} 
By adopting best practices for dataset documentation and attribution, the MIR community can demonstrate its commitment to ethical AI development and foster a more collaborative relationship with creators and rightholders.\footnote{For a philosophical perspective on the exploitative nature of AI art, see \citet{Goetze2024}.}

\subsection{Outlook and Future Directions}

Looking ahead, several areas warrant further exploration:

\textbf{Harmonization of Legal Frameworks:} 
The global nature of AI research and data use underscores the need for international dialogue on harmonizing legal frameworks for generative AI training. 
Policymakers, researchers, and industry stakeholders must collaborate to develop clear and consistent guidelines that balance innovation with the protection of intellectual property rights.
This paper should have made it clear that such guidelines can have a great impact on MIR research.
We therefore hope that this paper will encourage the MIR community to get more involved in the ongoing debate.

\textbf{Understanding Memorization and Developing Technical Solutions:}
A deeper understanding of memorization mechanisms in generative models is crucial for developing effective technical solutions. 
While factors like model size, data duplication, and context length influence memorization, the precise conditions and neural processes behind it remain poorly understood.
Future research should explore how associative memory behaviors emerge, examining internal representations and the interplay between model architecture and data characteristics.
Furthermore, continued research into mitigating memorization in generative models is essential. 
This includes developing architectures that minimize unintended recall, as well as tools for detecting and filtering memorized content in model outputs.
By advancing both the understanding of memorization and the development of technical countermeasures, researchers can create generative models that balance creativity with responsibility, minimizing legal risks while preserving innovation.

\textbf{Standardization of Dataset Documentation:} 
The MIR community can play a leading role in establishing standardized best practices for dataset documentation. 
This could involve the creation of open-source tools for metadata extraction, fingerprinting, and content identification, as well as the development of shared databases for training data attribution.

\textbf{Engagement with Creators and Rightholders:} 
Building trust and collaboration with creators and rightholders is critical for the sustainable development of generative AI. 
Initiatives such as public fingerprint databases and machine-readable opt-out mechanisms can empower creators to assert their rights while enabling researchers to comply with legal and ethical standards.
This could be supported by technical solutions developed within the MIR community.

\textbf{Community Action:} 
Given that the most egregious copyright violations often occur in commercial AI systems, the MIR community faces a choice: Should we accept sponsorship from companies that engage in massive infringement? Should we ban papers that use datasets lacking proper documentation? These are difficult questions, but they deserve open debate. By establishing clear ethical guidelines, the community can signal that it values integrity and will not tacitly endorse harmful practices.

\textbf{Ethical AI Development:} 
As generative AI continues to evolve, the MIR community must prioritize ethical considerations in its research and practices. 
This includes not only respecting copyright but also addressing broader issues such as bias, fairness, and the societal impact of AI-generated content.

\subsection{Final Thoughts}

The intersection of generative AI and copyright law presents both challenges and opportunities for the MIR community. 
By addressing the legal, technical, and ethical dimensions of AI training, researchers can contribute to the development of generative systems that are not only innovative but also respectful of the rights and contributions of creators. 
Ultimately, this approach will help to sustain human creators and thereby guarantee the future of human creativity as the basis for generative AI -- a goal that everybody should be able to subscribe to.
The path forward requires a collaborative effort, involving researchers, policymakers, rightholders, and industry stakeholders, to ensure that generative AI serves as a tool for creativity and progress rather than a source of conflict and uncertainty.

In conclusion, while the road ahead is complex, the MIR community is well-equipped to navigate these challenges and lead the way toward a future where generative AI and creativity coexist harmoniously. 
By embracing transparency, innovation, and ethical responsibility, we can unlock the full potential of generative AI while upholding the principles of fairness and respect that underpin creative expression.
As the MIR community, we can demonstrate that we deeply care about the art of music and the artists by making an effort to properly acknowledge the people whose data we use and giving them a chance to provide or withdraw consent.

In the long term, embedding principles of fair attribution into dataset creation and usage should become standard practice. 
One possible step forward is the inclusion of an ethics statement in research papers explicitly detailing dataset sources, artist consent mechanisms, and steps taken to ensure transparency. 
Such a shift would align with broader efforts to make AI development more accountable while reinforcing the music research community's dedication to both technological advancement and artistic integrity.

\section{Competing interests}
  
Parts of this article are based on an interdisciplinary study on generative AI and copyright law in Germany and the EU conducted by the authors in 2024 and funded by the German Authors' Rights Initiative (Initiative Urheberrecht).
The contents of the resulting report \citep{2024:dornis:urheberrec} as well as this article exclusively reflect the views and assessments of the authors.

\IfFileExists{\jobname.ent}{
   \theendnotes
}{
}

\section*{Acknowledgements}

The main points covered in this article were originally presented in a virtual special session talk by Sebastian Stober at the ISMIR 2024 conference. 
ChatGPT and Deepseek-R1 were used to turn the transcript of this talk into a first draft of this article as part of an experiment to assess the capabilities of these generative AI systems. 
This draft was subsequently altered significantly and edited thoroughly as well as extended by the authors to cover additional points and fully represent their views. 
The authors would like to express their gratitude to the reviewers for their constructive feedback that was highly valuable for improving the manuscript.


\bibliography{references}

\end{document}